\documentclass{revtex4}
\usepackage[T1]{fontenc}
\usepackage[latin1]{inputenc}
\usepackage[T1]{fontenc}
\usepackage[latin1]{inputenc}
\usepackage{float}
\usepackage{graphicx}

\begin{document}
\begin{flushright} JLAB-THY-07-716 \\
                   September 6, 2007
\end{flushright}
\begin{flushright}\vspace{2cm}
\end{flushright}

\title{Deeply Virtual Neutrino Scattering%
\footnote{Talk given at the Fifth International Workshop on Neutrino-Nucleus 
Interactions in the Few-GeV Region, FNAL, May 30 - June 3, 2007.%
          }}
\author{A. PSAKER}
\affiliation{Physics Department, Hampton University,\\
             Hampton, VA 23668, USA\\
             and\\
             Theory Group, Jefferson Laboratory,\\
             12000 Jefferson Avenue, Newport News, VA 23606, USA\\
             }

\begin{abstract}
\vspace{2cm}

We investigate the extension of the deeply virtual Compton scattering process into the weak interaction sector.

\vspace{5mm}

PACS number(s): 13.15.+g, 13.40.-f, 13.40.Gp, 13.60.-r, 13.60.Fz
\end{abstract}

\maketitle
\newpage

\section{Introduction}

Standard electromagnetic Compton scattering provides a unique tool for studying 
hadrons, which is one of the most fascinating frontiers of modern science. In this 
process the relevant Compton scattering amplitude probes the hadron structure by 
means of two quark electromagnetic currents. We argue that replacing one of the 
currents with the weak interaction current can promise a new insight.\\

The paper is organized as follows. In Sec.~\ref{handbag factorization} we briefly 
discuss the features of the handbag factorization scheme. We introduce a new set of 
phenomenological functions, known as generalized parton distributions (GPDs) 
\cite{Muller:1998fv,Ji:1996ek,Ji:1996nm,Radyushkin:1996nd,Radyushkin:1996ru,Radyushkin:1997ki}, 
and discuss some of their basic properties in Sec.~\ref{generalized parton 
distributions}. An application of the GPD formalism to the neutrino-induced deeply 
virtual Compton scattering in the kinematics relevant to future high-intensity 
neutrino experiments is given in Sec.~\ref{deeply virtual neutrino scattering}. 
The cross section results are presented in Sec.~\ref{results}. Finally, in 
Sec.~\ref{conclusions} we draw some conclusions and discuss future prospects.\\

Some of the formal results in this paper have appeared in preliminary reports in 
Refs.~\cite{Psaker:2004sf} and \cite{Psaker:2005wj}, whereas a comprehensive 
analysis of the weak neutral and weak charged current DVCS reactions in 
collaboration with W.~Melnitchouk and A.~Radyushkin has been presented in 
Ref.~\cite{Psaker:2006gj}. Neutrino scattering off nucleons for neutral and 
charged currents was also discussed in Refs.~\cite{Amore:2004ng} 
and \cite{Coriano:2004bk}, respectively.

\section{Handbag Factorization \label{handbag factorization}}

The underlying theory of strong interaction physics is the universally accepted 
nonAbelian gauge field theory known as quantum chromodynamics (QCD). The theory 
postulates that hadrons are composite objects made up of quarks, and that the color 
interaction between them is mediated by gluons as the gauge bosons. In principle, 
QCD embraces all phenomena of hadronic physics, in other words, the fundamental 
building blocks of hadrons (i.e. quarks and gluons) are known, and further the 
interactions between them are described by the well-established QCD Lagrangian. 
Nevertheless, knowing the first principles is not sufficient. The main difficulty is 
that QCD is formulated in terms of colored degrees-of-freedom, while the physical 
hadrons observed in the experiments are colorless. How to translate the information 
obtained from the experiments at the hadronic level into the language of quark and 
gluon fields has yet to be answered, and it represents a challenging task.\\

\begin{figure}[H]
\begin{center}
\includegraphics[%
  scale=0.5]{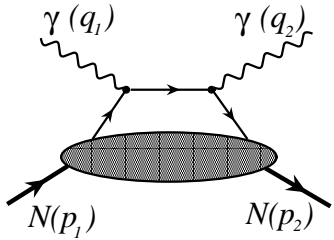}
\end{center}\caption{Handbag diagram. In the lowest QCD approximation, both photons 
couple to the same quark line through point-like vertices.}
\label{handbag}
\end{figure}

One way to tackle this problem is to consider projecting these fields onto the 
hadronic states, and then use different phenomenological functions (e.g. form 
factors, parton distribution functions (PDFs), distribution amplitudes and GPDs) to 
describe the resulting hadronic matrix elements. The key idea of this approach is 
the handbag factorization. According to this property, in the specific kinematical 
regime featuring the presence of a large invariant, such as the virtuality of the 
probe or the momentum transfer, the amplitude of the process splits into the hard 
(short-distance) part with only one active parton propagating along the light-cone 
and the soft (long-distance) part represented by the lower blob, see 
Fig.~\ref{handbag}. The asymptotically free nature of QCD allows us to compute the 
short-distance interactions by means of perturbation theory, whereas the 
nonperturbative stages of interactions are expressed in terms of well-defined 
quark-gluon operators taken on the light-cone, and being sandwiched between the 
hadronic states. These matrix elements, which accumulate the information about the 
long-distance dynamics of the process, are process independent nonperturbative 
objects. Thus they are parametrized in terms of universal phenomenological 
functions, and they can be measured with the help of different probes (e.g. photons 
and weak interaction bosons).\\

There are several situations, where the handbag contribution plays an essential 
role:
\begin{itemize}
\item Both initial and final photons are highly virtual and have equal space-like 
virtualities. This situation corresponds to the forward virtual Compton scattering 
amplitude. Its imaginary part, through the optical theorem, determines the structure 
functions of deeply inelastic lepton scattering (DIS).
\item The condition on photon virtualities may be relaxed in a sense that the 
initial photon is still far off-shell but the final photon is real and the invariant 
momentum transfer to the hadron is small. The situation corresponds to the 
nonforward virtual Compton scattering amplitude, which is accessible through 
processes such as deeply virtual Compton scattering (DVCS) and deeply virtual meson 
production.
\item The configuration, in which both photons in the initial and final states are 
real but the invariant momentum transfer is large. The physical process 
corresponding to this situation is known as wide-angle real Compton scattering.
\end{itemize}

\section{Generalized Parton Distributions \label{generalized parton distributions}}

The description of specific aspects of the hadron structure is provided by several 
different phenomenological functions. The well-known examples are the so-called old 
phenomenological functions such as form factors, PDFs and distribution amplitudes 
since they have been around for a long time. On the other hand, the concept of GPDs 
is new. These new functions, as hybrids of the old ones, provide the most complete 
information about the hadron structure. In fact, the old phenomenological functions 
are just the limiting case of GPDs. Like PDFs, GPDs encapsulate longitudinal 
momentum distributions of hadron's constituents, however, at the same time, like 
form factors, they also provide the information about their transverse coordinate 
distributions, and hence give a comprehensive three-dimensional snapshot of the 
substructure of the hadron. Moreover, the universality of GPDs enables one to 
develop a unified description of wide variety of different hard (i.e. light-cone 
dominated) processes, both inclusive and exclusive.\\

\begin{figure}[H]
\begin{center}
\includegraphics[%
  scale=0.5]{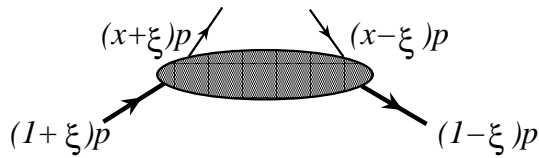}
\end{center}\caption{Symmetric partonic picture.}
\label{partonicpicture}
\end{figure}

In the symmetric partonic picture, see Fig.~\ref{partonicpicture}, we treat both the 
initial and final hadron momenta symmetrically by introducing the average hadron 
momentum $p\equiv\left(p_{1}+p_{2}\right)/2$. Here $x$ is the usual light-cone 
momentum fraction. In addition, another scaling variable is introduced, the so-
called skewness $\xi$. It is defined as the coefficient of the proportionality 
between the light-cone plus components of the momentum transfer, 
$r\equiv p_{1}-p_{2}$, and average hadron momentum, $\xi\equiv r^{+}/2p^{+}$, and it 
specifies the longitudinal momentum asymmetry. Clearly, in the skewed (nonforward) 
kinematics GPDs uncover much richer information about the hadron structure, which is 
not accessible in DIS, or in general in any inclusive process. By removing the 
parton with the light-cone momentum fraction $x+\xi$ from the hadron, and replacing 
it at some later point on the light-cone with the parton of fraction $x-\xi$, GPDs 
measure the coherence between the two parton momentum states in the hadron as well 
as their spin correlations. Finally, by considering two different hadrons in the 
initial and final states one can study flavor nondiagonal GPDs, for example, in the 
proton-to-neutron transition accessible through the exclusive charged pion 
electroproduction, the proton-to-$\Lambda$ transition in the kaon electroproduction 
or the nucleon-to-$\Delta$ transition.\\
Since $-1\leq x\leq1$, the momentum fractions $x\pm\xi$ of the active partons can be 
either positive or negative, and for that reason GPDs have three distinct regions. 
When $\xi\leq x\leq1$ both partons represent quarks, while for $-1\leq x\leq-\xi$ 
they are both antiquarks. In these two regions GPDs are just a generalization of 
PDFs. In the central region, $-\xi\leq x\leq\xi$, which is often referred to 
as the mesonic region, the partons represent the quark-antiquark pair. Here GPDs 
behave like meson distribution amplitudes.\\

At the leading twist-2 level, the hadron structure information can be parametrized 
in terms of two unpolarized and two polarized GPDs denoted by $H$, $E$ and 
$\widetilde{H}$, $\widetilde{E}$, respectively. They are functions of three 
variables $\left(x,\xi,t\right)$, where $t=r^2$ is the invatiant momentum transfer,  
and further they are defined for each quark flavor $f$. GPDs have interesting 
properties linking them to PDFs and form factors. In the forward limit, 
$p_{1}=p_{2}$, $r=0$, $\xi=0$, $t=0$, they reduce to PDFs. In particular, $H_{f}$ 
and $\widetilde{H}_{f}$ coincide with the quark density distribution 
$f_{N}\left(x\right)$ and the quark helicity distribution 
$\Delta f_{N}\left(x\right)$. We write the so-called reduction formulas,
\begin{equation}
H_{f}\left(x,0,0\right)=\left\{ \begin{array}{cc}
f_{N}\left(x\right) & x>0\\
-\bar{f}_{N}\left(-x\right) & x<0\\
\end{array}\right.
\label{eq:forwardlimit1}
\end{equation}
and
\begin{equation}
\widetilde{H}_{f}\left(x,0,0\right)=\left\{ \begin{array}{cc}
\Delta f_{N}\left(x\right) & x>0\\
\Delta\bar{f}_{N}\left(-x\right) & x<0,\\
\end{array}\right.
\label{eq:forwardlimit2}
\end{equation}
while $E_{f}$ and $\widetilde{E}_{f}$ have no connections to PDFs. They are always 
accompanied with the momentum transfer, and hence simply not visible in DIS. 
Even though they have no analogue in the forward limit their limits do exist. In the 
local limit GPDs reduce to the standard nucleon elastic form factors (i.e. Dirac, 
Pauli, axial and pseudoscalar form factors),
\begin{eqnarray}
\int_{-1}^{1}dx\; H_{f}\left(x,\xi,t\right)=F_{1f}\left(t\right) & , & 
\int_{-1}^{1}dx\; E_{f}\left(x,\xi,t\right)=F_{2f}\left(t\right),\\
\int_{-1}^{1}dx\;\widetilde{H}_{f}\left(x,\xi,t\right)=g_{Af}\left(t\right) & , & 
\int_{-1}^{1}dx\;\widetilde{E}_{f}\left(x,\xi,t\right)=g_{Pf}\left(t\right).
\label{eq:sumrules}
\end{eqnarray}
We call these relations the sum rules.\\

GPDs in general are also relevant for the hadron spin structure. In terms of GPDs 
their second moments at $t=0$ give the quark total angular momentum,
\begin{eqnarray}
J_{q} & = & \frac{1}{2}\sum_{f}\int_{-1}^{1}dx\; x
\left[H_{f}\left(x,\xi,t=0\right)+E_{f}\left(x,\xi,t=0\right)\right].
\label{eq:quarkangularmomentum}
\end{eqnarray}
On the other hand, by decomposing $J_{q}$ into the quark intrinsic spin 
$\Delta\Sigma$ (measured through polarized DIS) and quark orbital angular momentum 
$L_{q}$, $J_{q}=\Delta\Sigma/2+L_{q}$, we can access $L_{q}$. Finally, since the 
total nucleon spin comes from quarks and gluons, $1/2=J_{q}+J_{g}$, we can extract 
the total angular momentum $J_{g}$ carried by gluons.

\section{Deeply Virtual Neutrino Scattering \label{deeply virtual neutrino scattering}}

In recent years, significant effort was made to study GPDs through measurements of 
hard exclusive electroproduction processes. The simplest process in this respect is 
the standard electromagnetic DVCS process. Complementary to DVCS different 
combinations of quark flavors can be accessed by utilizing the weak interaction 
current in neutrino-induced DVCS (or, alternatively deeply virtual neutrino 
scattering - DVNS), namely
\begin{eqnarray}
\nu\left(k\right)N\left(p_{1}\right) & \longrightarrow &
e^{-}\left(k'\right)N'
\left(p_{2}\right)\gamma\left(q_{2}\right)
\label{eq:CCreaction}
\end{eqnarray}
for the charged current, and
\begin{eqnarray}
\nu\left(k\right)N\left(p_{1}\right) & \longrightarrow &
\nu\left(k'\right)N
\left(p_{2}\right)\gamma\left(q_{2}\right)
\label{eq:NCreaction}
\end{eqnarray}
for neutral current reactions. The presence of the axial part of the $V-A$ 
interaction enables to probe a different set of GPDs, i.e. the charge conjugation 
odd combinations of GPDs as well as charge conjugation even combinations and hence 
independently measure both the valence and sea content of GPDs. Furthermore, one 
deals with a different flavor decomposition, in particular, less enhancement from 
$u$-quarks. An additional feature of the DVNS process is the study of GPDs that are 
not diagonal in quark flavors, such as those associated with the neutron-to-proton 
transition.\\

\begin{figure}[H]
\begin{center}
\includegraphics[%
  scale=0.7]{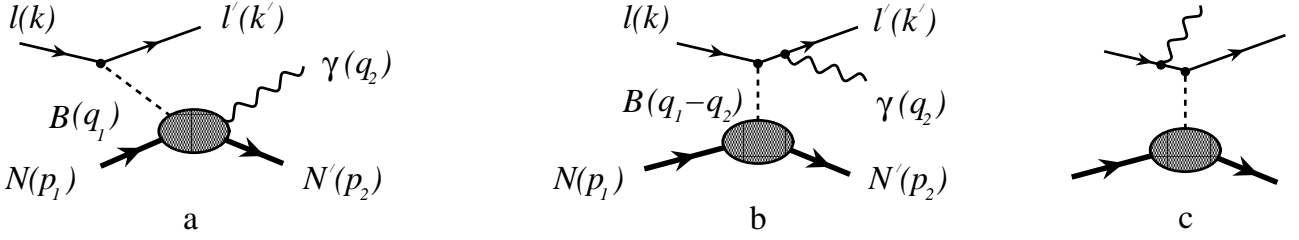}
\end{center}\caption{DVCS (a) and Bethe-Heitler (b and c) diagrams contributing to 
the leptoproduction of a real photon.}
\label{dvns}
\end{figure}

In the most general case, there are three relevant diagrams for a DVCS-like process, 
see Fig.~\ref{dvns}. The nucleon blob with the boson leg, which can be either a 
virtual photon or weak interaction boson, and photon leg represents the virtual 
Compton scattering amplitude. This diagram is referred to as the DVCS diagram or the 
Compton contribution. The real photon can also be emitted by a lepton as depicted in 
the remaining two diagrams. They illustrate the Bethe-Heitler process, where the 
lower blob with one boson leg stands for the electroweak form factor, while the 
upper part in each diagram can be exactly calculable in QED.\\

\begin{figure}[H]
\begin{center}
\includegraphics[%
  scale=0.6]{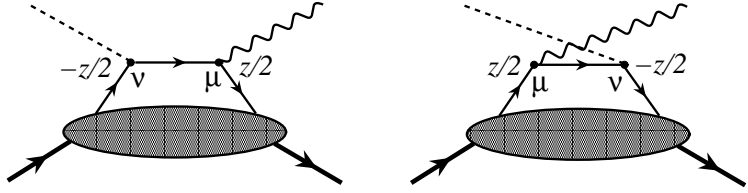}
\end{center}\caption{The $s$- and $u$-channel handbag diagrams for the virtual 
Compton scattering amplitude.}
\label{dvnshandbag}
\end{figure}

In the Bjorken regime, in which the virtuality of the weak boson $-q^{2}_{1}$ and 
the invariant mass of the weak boson-nucleon system, 
$s\equiv\left(p_{1}+q_{1}\right)^{2}$, are both sufficiently large while the Bjorken 
ratio $x_{B}\equiv-q_{1}^{2}/\left[2\left(p_{1}\cdot q_{1}\right)\right]$ is kept 
finite, the behavior of the virtual Compton scattering amplitude is dominated by the 
light-like distances. The dominant light-cone singularities are represented by two 
($s$- and $u$-channel) handbag diagrams, see Fig.~\ref{dvnshandbag}, in which the 
hard quark propagator is convoluted with GPDs. In addition, by keeping the invariant 
momentum transfer to the nucleon as small as possible we arrive at the so-called 
DVCS kinematics, $s>-q_{1}^{2}\gg-t$. Accordingly, the $t$-dependence enters only in 
the soft part of the amplitude.\\

In our case in terms of symmetric momentum variables, 
$p\equiv\left(p_{1}+p_{2}\right)/2$ and $q\equiv\left(q_{1}+q_{2}\right)/2$, the 
virtual Compton scattering amplitude is given by the Fourier transform of the 
correlation function of the electromagnetic and weak (neutral or charged) 
interaction currents,
\begin{eqnarray}
\mathsf{\mathcal{T}}_{W}^{\mu\nu} & = & i\int d^{4}z\; e^{iq\cdot z}
\left\langle N'\left(p-r/2,s_{2}\right)\right|T\left\{J_{EM}^{\mu}
\left(z/2\right)J_{W}^{\nu}\left(-z/2\right)\right\} \left|N
\left(p+r/2,s_{1}\right)\right\rangle\ .
\label{eq:reducedamplitude}
\end{eqnarray}
The weak current couples to the quark current through two types of vertices, 
$qqZ^{0}$ and $qqW^{\pm}$, and hence the quark fields at coordinates $\pm z$ can 
either have the same or different flavor quantum numbers. For that reason, we treat 
these two cases separately. The method employed to study the amplitude 
(\ref{eq:reducedamplitude}) in the DVCS kinematics is based on the nonlocal 
light-cone expansion of the time-ordered product of currents in terms of QCD bilocal 
operators in coordinate space \cite{Balitsky:1987bk},
\begin{eqnarray}
iT\left\{ J_{EM}^{\mu}\left(z/2\right)J_{WN}^{\nu}
\left(-z/2\right)\right\}  
& = & -\frac{z_{\rho}}{4\pi^{2}z^{4}}
\sum_{f}Q_{f}\left\{ c_{V}^{f}\left[s^{\mu\rho\nu\eta}
\mathcal{O}_{\eta}^{f-}\left(z\left|0\right.\right)-i
\epsilon^{\mu\rho\nu\eta}\mathcal{O}_{5\eta}^{f+}
\left(z\left|0\right.\right)\right]\right.\nonumber \\
&  & \left.-\ c_{A}^{f}\left[s^{\mu\rho\nu\eta}\mathcal{O}_{5\eta}^{f-}
\left(z\left|0\right.\right)-i\epsilon^{\mu\rho\nu\eta}
\mathcal{O}_{\eta}^{f+}\left(z\left|0\right.\right)\right]\right\}\ ,
\label{eq:weakneutralexpansion2}
\end{eqnarray}
where $Q_{f}$ denotes the electric charge of the quark with flavor $f$ (in units of 
$\left|e\right|$), $c_{V}^{f}$ and $c_{A}^{f}$ are the corresponding weak vector and 
axial vector charges,
$s^{\mu\rho\nu\eta}\equiv g^{\mu\rho}g^{\nu\eta}+g^{\mu\eta}g^{\rho\nu}-
g^{\mu\nu}g^{\rho\eta}$ is the symmetric tensor and $\epsilon^{\mu\rho\nu\eta}$ the 
antisymmetric tensor in Lorentz indices ($\mu,\nu$), and further
\begin{eqnarray}
\label{eq:uncontractedstringoperators1}
\mathcal{O}_{\eta}^{f\pm}\left(z\left|0\right.\right) & \equiv & 
\left[\bar{\psi}_{f}\left(z/2\right)\gamma_{\eta}\psi_{f}
\left(-z/2\right)\pm\left(z\rightarrow-z\right)\right]\ , \\
\mathcal{O}_{5\eta}^{f\pm}\left(z\left|0\right.\right) & \equiv & 
\left[\bar{\psi}_{f}\left(z/2\right)\gamma_{\eta}
\gamma_{5}\psi_{f}\left(-z/2\right)\pm
\left(z\rightarrow-z\right)\right]\ ,
\label{eq:uncontractedstringoperators2}
\end{eqnarray}
are the vector and axial vector bilocal operators, respectively. Unlike the weak 
neutral current sector explicily presented here, the expansion for the weak charged 
and electromagnetic currents is a bit more complicated due to the fact that 
$W^{\pm}$ carries an electric charge. Thus we end up with different nucleons in the 
initial and final states and accordingly, the bilocal operators are accompanied with 
different electric charges as well as flavors.\\

Next we isolate the twist-2 part of these operators, and sandwich it between the 
initial and final nucleon states. To construct the parametrization for nonforward 
matrix elements on the light-cone, we use a spectral representation in terms of 
eight GPDs, which have well-defined symmetry properties with respect to $x$. In 
particular, in case of the weak neutral current we have
\begin{eqnarray}
\label{eq:wnparametrization1}
\left\langle N\left(p_{2},s_{2}\right)\right|\mathcal{O}^{f\pm}
\left(z\left|0\right.\right)
\left|N\left(p_{1},s_{1}\right)\right\rangle_{z^2=0}  
& = & \bar{u}\left(p_{2},s_{2}
\right)\not\! zu\left(p_{1},s_{1}\right)
\int_{-1}^{1}dx\; e^{ix p \cdot z}H_{f}^{\pm}
\left(x,\xi,t\right) \nonumber \\
&  & +\bar{u}\left(p_{2},s_{2}\right)
\frac{\left(\not\! z\not\! r-\not\! r\not\! z\right)}{4M}u
\left(p_{1},s_{1}\right)
\int_{-1}^{1}dx\; e^{ix p \cdot z}E_{f}^{\pm}
\left(x,\xi,t\right)\ , \\
\left\langle N\left(p_{2},s_{2}\right)\right|\mathcal{O}_{5}^{f\pm}
\left(z\left|0\right.\right)\left|N\left(p_{1},s_{1}\right)\right
\rangle_{z^2=0}  & = & \bar{u}\left(p_{2},s_{2}\right)\not\! z\gamma_{5}u
\left(p_{1},s_{1}\right)\int_{-1}^{1}dx\; e^{ix p \cdot z}
\widetilde{H}_{f}^{\mp}\left(x,\xi,t\right)\nonumber \\
&  & -\bar{u}\left(p_{2},s_{2}\right)\frac{\left(r\cdot z\right)}{2M}
\gamma_{5}u\left(p_{1},s_{1}\right)
\int_{-1}^{1}dx\; e^{ix p \cdot z}
\widetilde{E}_{f}^{\mp}\left(x,\xi,t\right)\ .
\label{eq:wnparametrization2}
\end{eqnarray}
In contrast to the electromagnetic DVCS process, which measures only the {\em plus} 
GPDs (i.e. the sum of quark and antiquark distributions), the DVNS process gives 
also access to the {\em minus} distributions (i.e. the valence configuration) 
associated with the axial part of the $V-A$ interaction. Moreover, it's important to 
note that since the matrix elements with the weak charged interaction current 
involve different flavor combinations the corresponding GPDs, which parametrize 
these matrix elements, are nondiagonal in quark flavor.\\

The final expression for the weak neutral leading twist-2 Compton scattering 
amplitude reads
\begin{eqnarray}
\mathcal{T}_{WN}^{\mu\nu} & = & 
-\frac{1}{4\left(p\cdot q\right)}
\left\{ \left[
\frac{1}{\left(p\cdot q_{2}\right)}
\left(p^{\mu}q_{2}^{\nu}+p^{\nu}q_{2}^{\mu}\right)-g^{\mu\nu}
\right]\right.\nonumber \\
&  & \times\left[\mathcal{H}_{WN}^{+}\left(\xi,t\right)\bar{u}
\left(p_{2},s_{2}\right)
\not\! q_{2}u\left(p_{1},s_{1}\right)+
\mathcal{E}_{WN}^{+}\left(\xi,t\right)
\bar{u}\left(p_{2},s_{2}\right)
\frac{\left(\not\! q_{2}
\not\! r-\not\! r\not\! q_{2}\right)}{4M}u\left(p_{1},s_{1}\right)
\right.\nonumber \\
&  & \hspace*{0.5cm}
     \left.-\widetilde{\mathcal{H}}_{WN}^{-}\left(\xi,t\right)
\bar{u}\left(p_{2},s_{2}\right)\not\! q_{2}\gamma_{5}u
\left(p_{1},s_{1}\right)+
\widetilde{\mathcal{E}}_{WN}^{-}\left(\xi,t\right)
\frac{\left(q_{2}\cdot r\right)}{2M}
\bar{u}\left(p_{2},s_{2}\right)
\gamma_{5}u\left(p_{1},s_{1}\right)\right]\nonumber \\
&  & +\left[\frac{1}{\left(p\cdot q_{2}\right)}i
\epsilon^{\mu\nu\rho\eta}q_{2\rho}p_{\eta}
\right]\nonumber \\
&  & \times\left[\widetilde{\mathcal{H}}_{WN}^{+}\left(\xi,t\right)
\bar{u}\left(p_{2},s_{2}\right)\not\! q_{2}
\gamma_{5}u\left(p_{1},s_{1}\right)-\widetilde{\mathcal{E}}_{WN}^{+}
\left(\xi,t\right)
\frac{\left(q_{2}\cdot r\right)}{2M}
\bar{u}\left(p_{2},s_{2}\right)\gamma_{5}u\left(p_{1},s_{1}\right)
\right.\nonumber \\
&  & \hspace*{0.5cm}
     \left.\left.-\mathcal{H}_{WN}^{\mathit{-}}\left(\xi,t\right)
\bar{u}\left(p_{2},s_{2}
\right)\not\! q_{2}u\left(p_{1},s_{1}\right)-
\mathcal{E}_{WN}^{\mathit{-}}\left(\xi,t\right)
\bar{u}\left(p_{2},s_{2}\right)
\frac{\left(\not\! q_{2}\not\! r-\not\! r\not\! q_{2}
\right)}{4M}u\left(p_{1},s_{1}\right)
\right]\right\}\ ,
\label{eq:compactweakneutralamplitude}
\end{eqnarray}
where
\begin{eqnarray}
\mathcal{H}_{WN}^{+\left(-\right)}\left(\xi,t\right) & 
\equiv & \sum_{f}Q_{f}c_{V\left(A\right)}^{f}\int_{-1}^{1}
\frac{dx}{\left(x-\xi+i0\right)}H_{f}^{+\left(-\right)}
\left(x,\xi,t\right) \nonumber \\
& = & \sum_{f}Q_{f}c_{V\left(A\right)}^{f}
\int_{-1}^{1}dx\; H_{f}\left(x,\xi,t\right)
\left(\frac{1}{x-\xi+i0}\pm\frac{1}{x+\xi}\right)\;, \\
\mathcal{E}_{WN}^{+\left(-\right)}\left(\xi,t\right) & 
\equiv & \sum_{f}Q_{f}c_{V\left(A\right)}^{f}\int_{-1}^{1}
\frac{dx}{\left(x-\xi+i0\right)}E_{f}^{+\left(-\right)}
\left(x,\xi,t\right) \nonumber \\
& = & \sum_{f}Q_{f}c_{V\left(A\right)}^{f}
\int_{-1}^{1}dx\; E_{f}\left(x,\xi,t\right)
\left(\frac{1}{x-\xi+i0}\pm\frac{1}{x+\xi}
\right)\;, \\
\widetilde{\mathcal{H}}_{WN}^{+\left(-\right)}
\left(\xi,t\right) & \equiv & 
\sum_{f}Q_{f}c_{V\left(A\right)}^{f}\int_{-1}^{1}
\frac{dx}{\left(x-\xi+i0\right)}\widetilde{H}_{f}^{+\left(-\right)}
\left(x,\xi,t\right) \nonumber \\
& = & \sum_{f}Q_{f}c_{V\left(A\right)}^{f}
\int_{-1}^{1}dx\;\widetilde{H}_{f}\left(x,\xi,t\right)
\left(\frac{1}{x-\xi+i0}\mp\frac{1}{x+\xi}\right)\;, \\
\widetilde{\mathcal{E}}_{WN}^{+\left(-\right)}\left(\xi,t\right) & 
\equiv & \sum_{f}Q_{f}c_{V\left(A\right)}^{f}\int_{-1}^{1}
\frac{dx}{\left(x-\xi+i0\right)}\widetilde{E}_{f}^{+\left(-\right)}
\left(x,\xi,t\right) \nonumber \\
& = & \sum_{f}Q_{f}c_{V\left(A\right)}^{f}
\int_{-1}^{1}dx\;\widetilde{E}_{f}\left(x,\xi,t\right)
\left(\frac{1}{x-\xi+i0}\mp\frac{1}{x+\xi}\right)\;,
\label{eq:integralsofGPDsweakneutralcurrent}
\end{eqnarray}
are the so-called Compton form factors given by the integrals of GPDs. The amplitude 
(\ref {eq:compactweakneutralamplitude}) has both real and imaginary parts. The real 
part is obtained by the principal value prescription, whereas the imaginary part 
constrains evaluating GPDs at the specific point $x=\pm\xi$.\\

\begin{figure}[H]
\begin{center}
\includegraphics[%
  scale=0.6]{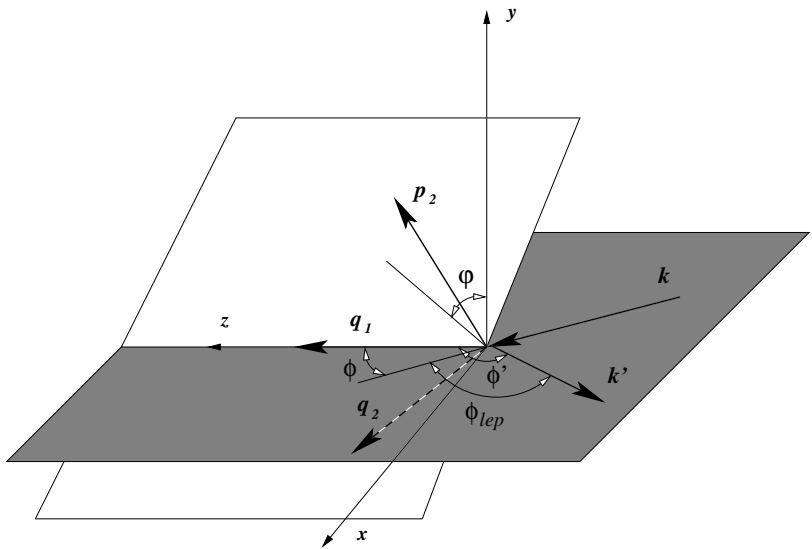}
\end{center}\caption{The kinematics of the generalized DVCS process in the target 
rest frame. The lepton momenta $k$ and $k'$ define the lepton scattering plane and 
the momenta of the final photon and outgoing nucleon define the nucleon scattering 
plane.}
\label{kinematics}
\end{figure}

\begin{figure}[H]
\begin{center}
\includegraphics[%
  scale=1.0]{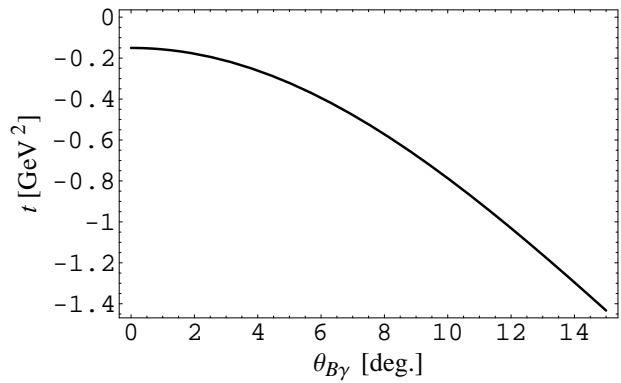}
\end{center}\caption{Invariant momentum transfer as a function of the scattering 
angle between the incoming virtual weak boson and outgoing real photon.}
\label{t}
\end{figure}

In the laboratory frame as the target rest frame we choose a coordinate system, in 
which the weak boson four-momentum has no transverse components, see 
Fig.~\ref{kinematics}. The differential cross section is given by
\begin{eqnarray}
\frac{d^{4}\sigma}{dx_{B}dQ_{1}^{2}dtd\varphi} & = & 
\frac{1}{32}\frac{1}{\left(2\pi\right)^{4}}
\frac{x_{B}y^{2}}{Q_{1}^{4}}
\frac{1}{\sqrt{1+4x_{B}^{2}M^{2}/Q_{1}^{2}}}
\left|{\textrm{T}}\right|^{2}\,\,,
\label{eq:generaldiffcrosswithQ1squared}
\end{eqnarray}
where the invariant matrix element is the sum of both the Compton and Bethe-Heitler 
contributions, $\mathrm{T}=\mathrm{T}_{C}+\mathrm{T}_{BH}$. Furthermore, we require 
the following:
\begin{itemize}
\item   The energy of the incoming lepton beam is fixed at 
	$\omega=20\;\mathrm{GeV}$.
\item   The invariant mass squared of the weak boson-nucleon
	system should be above the resonance region, 
	$s \geq 4\;\mathrm{GeV}^{2}$.
\item   The virtuality of the incoming boson has to be large enough
	to ensure light-cone dominance of the scattering process,
	$Q_{1}^{2}\geq 2.5\;\mathrm{GeV}^{2}$.
\item   The momentum transfer squared should be as small as possible,
	e.g. $0.1\;\mathrm{GeV^{2}}\leq\left|t\right|\leq0.2\;
	\mathrm{GeV^{2}}$.
\end{itemize}
We pick one kinematical point, in particular with the parameters 
$\omega=20\;\mathrm{GeV}$, $Q_{1}^{2}=2.5\; \mathrm{GeV}^{2}$, 
$x_{B}=0.35$ and $y=0.19$, and plot the invariant momentum transfer as a function of 
the angle $\theta_{B\gamma}$ between the incoming virtual weak boson and outgoing 
real photon, see Fig.~\ref{t}. The values vary from $-0.15\;\mathrm{GeV}^{2}$ at 
zero degrees up to $-1\;\mathrm{GeV}^{2}$ at $\theta_{B\gamma}=12^\circ$. Finally, 
we set the angle between the two planes to zero. In the so-called in-plane 
kinematics the polar angles of both incoming and scattered leptons are also fixed to 
$\phi=20.2^\circ$ and $\phi'=25.3^\circ$, respectively.

We use a simple model for nucleon GPDs 
\cite{Guichon:1998xv,Radyushkin:1998rt,Belitsky:2001ns,Carlitz:1976in,Goshtasbpour:1995eh}  
with the following assumptions: the sea quark contribution is neglibile and hence 
the plus and minus distributions coincide; the $t$-dependence of GPDs, which is 
driven by the corresponding nucleon form factors, factorizes from the dependence on 
the other two scaling variables, $x$ and $\xi$; the $\xi$-dependence of GPDs appears 
only in the $\widetilde{E}_{f}$ distribution.\\

In the weak neutral case we consider a neutrino-induced DVCS on a proton target 
through the exchange of $Z^{0}$. There is no Bethe-Heitler contamination (the photon 
cannot be emitted from the neutrino leg) and for that reason, we measure the pure 
Compton contribution,
\begin{eqnarray}
\mathrm{T}_{\nu p} & = & \sqrt{2}\left|e\right|G_{F}\bar{u}
\left(k'\right)\gamma_{\nu}
\left(1-\gamma_{5}\right)u(k)\epsilon_{\mu}^{*}
\left(q_{2}\right)\mathcal{T}_{WN}^{\mu\nu}\ .
\label{eq:Tmatrixneutrinoproton2}
\end{eqnarray}
In the weak charged current sector we examine scattering off neutrinos, which 
convert into muons by emission of $W^{+}$, on a neutron target. Here in addition to 
the Compton part,
\begin{eqnarray}
\mathrm{T}_{C\nu n} & = & \sqrt{2}\left|e\right|G_{F}\bar{u}
\left(k'\right)\gamma_{\nu}
\left(1-\gamma_{5}\right)u\left(k\right)\epsilon_{\mu}^{*}
\left(q_{2}\right)
\mathcal{T}_{WC}^{\mu\nu}\ ,
\label{eq:TmatrixCWC}
\end{eqnarray}
the Bethe-Heitler contribution is present,
\begin{eqnarray}
\mathrm{T}_{BH\nu n} & = & \sqrt{2}\left|e\right|G_{F}
\epsilon_{\mu}^{*}\left(q_{2}\right)
\bar{u}\left(k'\right)\left[\frac{\gamma^{\mu}
\left(\not\! k'+\not\! q_{2}\right)\gamma^{\nu}
\left(1-\gamma_{5}\right)}{\left(k'+q_{2}\right)^{2}}
\right]u\left(k\right)\left
\langle p\left(p_{2},s_{2}\right)
\right|J_{\nu}^{CC}\left(0\right)
\left|n\left(p_{1},s_{1}\right)\right\rangle\ ,
\label{eq:TmatrixBHcharged}
\end{eqnarray}
however, it's given only by diagram (b) of Fig.~\ref{dvns}. Moreover, the isospin 
symmetry is used to express the nucleon matrix elements that are nondiagonal in 
quark flavor in terms of flavor diagonal matrix elements \cite{Mankiewicz:1997aa}.

\section{Results \label{results}}

\begin{figure}[H]
\begin{center}
\includegraphics[%
  scale=1.0]{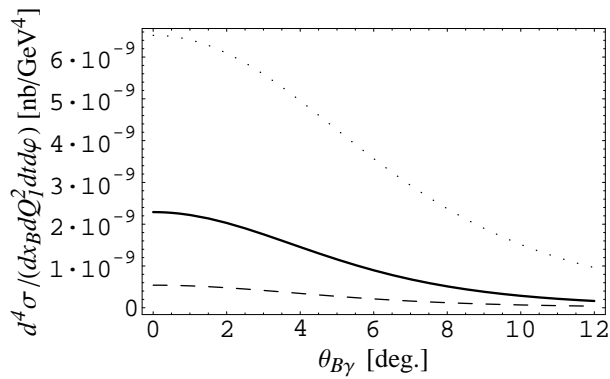}
\end{center}\caption{Compton cross sections for the weak neutral DVCS with the 
neutrino beam (solid line), electron beam (dashed line), and weak charged DVCS with 
the neutrino beam (dotted line).}
\label{weakDVCSCompton}
\end{figure}

\begin{figure}[H]
\begin{center}
\includegraphics[%
  scale=1.0]{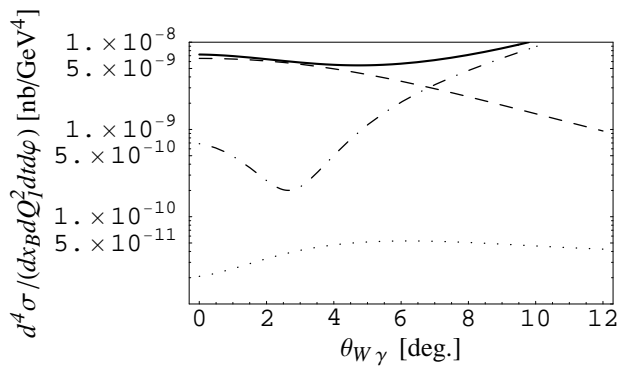}
\end{center}\caption{Compton (dashed line), Bethe-Heitler (dotted line), magnitude 
of interference (dashed-dotted line), and total (solid line) cross sections for 
weak charged DVCS.}
\label{wctotal}
\end{figure}

\begin{figure}[H]
\begin{center}
\includegraphics[%
  scale=1.0]{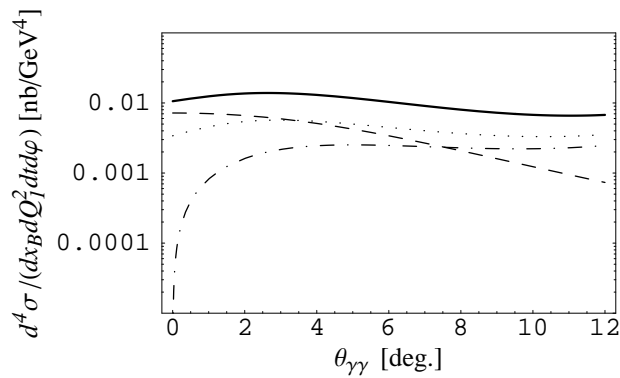}
\end{center}\caption{As in Fig.~\ref{wctotal} but for electromagnetic DVCS.}
\label{emtotal}
\end{figure}

\begin{figure}[H]
\begin{center}
\includegraphics[%
  scale=1.0]{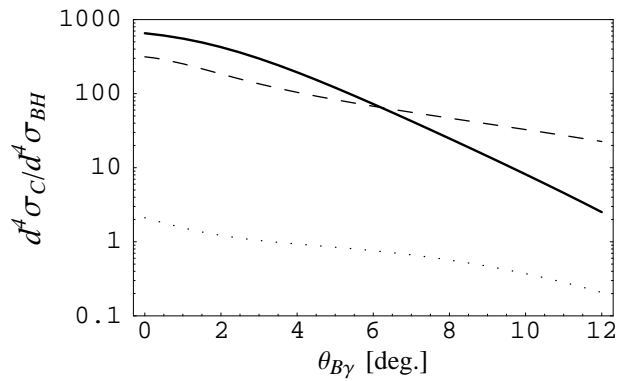}
\end{center}\caption{Ratio of Compton and Bethe-Heitler cross section for weak 
neutral DVCS with the electron beam (solid line), weak charged DVCS with the 
neutrino beam (dashed line), and electromagnetic DVCS (dotted line).}
\label{ratio}
\end{figure}

We present the cross section results for $\theta_{B\gamma}\leq12^\circ$, which 
corresponds to taking the invariant momentum transfer $-t < -1\;\mathrm{GeV}^{2}$. 
The charged current Compton cross section is larger than for the neutral current 
interaction, see Fig.~\ref{weakDVCSCompton}. We compare the magnitude of the 
Compton contribution relative to the corresponding Bethe-Heitler background by 
plotting both cross sections together on a logarithmic scale, see 
Fig.~\ref{wctotal}. The Compton cross section is significantly larger than the 
Bethe-Heitler signal. Note also that for the Bethe-Heitler cross section one should 
expect a pole at the scattering angle $\theta_{B\gamma}=25.3^\circ$ when the 
scattered muon is collinear with outgoing real photon. On the other hand, 
in the electromagnetic current case the Bethe-Heitler contribution is larger than 
the Compton for $\theta_{B\gamma}>3^\circ$, see Fig.~\ref{emtotal}. Finally, we plot 
the ratio of Compton versus Bethe-Heitler cross sections for both weak charged and 
electromagnetic DVCS. In the forward direction (i.e. for 
$\theta_{B\gamma}\leq4^\circ$) the Bethe-Heitler cross section is strongly supressed 
by a factor more than 100 compared to the Compton contribution for the weak charged 
interaction current, whereas in the electromagnetic current case the ratio between 
the cross sections remains of order unity, see Fig.~\ref{ratio}. 

\section{Summary and Outlook \label{conclusions}}

GPDs provide the most complete and unified description of the internal quark-gluon 
structure of hadrons, which can be probed through wide variety of both inclusive and 
exclusive hard processes.\\

DVNS is an important tool to complement the study of GPDs in the more familiar 
electro-induced DVCS. In addition to different combinations and flavor decomposition 
of GPDs, providing a direct measurement of their valence and sea contents, the 
process enables one to access the distributions that are nondiagonal in quark 
flavor.\\

We have computed the twist-2 Compton scattering amplitudes for both weak neutral and 
weak charged interaction currents by means of the light-cone expansion of the 
current product in coordinate space. Using a simple model for the nucleon GPDs, 
which only includes the valence quark contribution, we gave prediction for cross 
sections in the kinematics relevant to future high-intensity neutrino experiments.\\

Unlike the standard electromagnetic DVCS process, we find that at small scattering 
angles the Compton contribution is enhanced relative to the corresponding 
Bethe-Heitler contribution. Thus a contamination from the Bethe-Heitler background 
is less of a problem when extracting the weak DVCS signal.\\

In the future we plan to use more realistic model for nucleon GPDs, which also 
includes sea quark effects, elaborate separately contributions from the plus and 
minus distributions, and include the twist-3 terms in order to apply the formalism 
at moderate energies. 

\begin{acknowledgments}

Authored by Jefferson Science Associates, LLC under U.S. DOE Contract
No. DE-AC05-06OR23177. The U.S. Government retains a non-exclusive,
paid-up, irrevocable, world-wide license to publish or reproduce this
manuscript for U.S. Government purposes. 

\end{acknowledgments}


\end{document}